\documentclass{aa}  
\usepackage{graphicx}
\usepackage{natbib}
\usepackage{txfonts}
\begin{document}
 \title{Neon and oxygen in low activity stars: towards a coronal unification with the Sun}


   \author{J. Robrade
          \inst{1}
          \and
          J.H.M.M. Schmitt\inst{1}
\and
          F. Favata\inst{2}
          }


        \institute{Universit\"at Hamburg, Hamburger Sternwarte, Gojenbergsweg 112, D-21029 Hamburg, Germany\\
       \email{jrobrade@hs.uni-hamburg.de}
 \and
ESA - Planning and Community Coordination Office, 8-10 rue Mario Nikis, 75738 Paris Cedex 15, France
             }

   \date{Received 29 February 2008 ; accepted 27 May 2008}

 
  \abstract
{}
   {The disagreement between helioseismology and a recent downward revision of solar abundances has resulted in a controversy about the true neon abundance of the Sun and other stars.
We study the coronal Ne/O abundance ratios of nearby stars with modest activity levels and investigate a possible peculiarity of the Sun among the stellar population in the 
solar neighborhood.}
   {We used XMM-Newton and Chandra data from a sample of weakly and moderately active stars with log\,L$_{\rm X}$/L$_{\rm bol} \approx -5\,...-7 $ to investigate high-resolution X-ray spectra
to determine their coronal Ne/O abundance ratio. We applied two linear combinations of strong emission lines 
from neon and oxygen, as well as  a global-fitting method for each dataset, and crosschecked the derived results.}
   {The sample stars show a correlation between their Ne/O ratio and stellar activity in the sense that stars with a higher activity level show a higher Ne/O ratio. 
We find that the Ne/O abundance ratio decreases in our sample from values of Ne/O\,$\approx$\,0.4 down to Ne/O\,$\approx$\,0.2\,--\,0.25, suggesting that ratios
similar to 'classical' solar values, i.e. Ne/O\,$\approx0.2$, are rather common for low activity stars. A significantly enhanced
neon abundance as the solution to the solar modeling problem seems unlikely.}
   {From the coronal Ne/O abundance ratios, we find no indications of a peculiar position of the Sun among other stars.
The solar behavior appears to be rather typical of low activity stars.}

   \maketitle
%

\section{Introduction}

The chemical composition of the Sun is used as a reference frame throughout astronomy and therefore the precise determination 
of the solar abundances is of particular interest. 
Beside indirect methods utilizing solar wind particles and meteorites, the measurement of photospheric lines
is used as the prime method to derive the chemical composition of the Sun. 
Making use of more and more refined measurements and modeling, the chemical composition of the Sun 'evolved' throughout the decades, however
abundance compilations like the ones of \cite{angr} and more recently by \cite{grsa} have become widely accepted.
\cite{asplund} proposed the most recent revision of the solar abundance scale based on a new 3\,D hydrodynamic modeling of the solar atmosphere,
resulting in approximately 30\% lower abundances of many light elements including in particular C, N, O and Ne. 
While this new solar chemical composition provides a better agreement with e.g. measurements of the local ISM, 
on the solar side severe discrepancies appeared \citep{basu04, tur04,bah05}.
The light elements are an important source of opacity in the Sun and as a consequence the almost perfect agreement between helioseismology and models of
the solar interior based on the abundance values of \cite{grsa} is seriously disturbed. 
The disagreement is far beyond measurement errors and estimated modeling uncertainties and therefore a solution
of this problem is highly desired.

One proposal to reconcile the new abundances with helioseismology 
suggests an enhancement of the solar neon abundance \citep{ant05}. In contrast to C, N, O, Ne has no strong photospheric lines and its abundance
is determined from emission lines originating in the transition region and the corona or through composition measurements of solar wind particles.
However, both methods do not necessarily trace the true solar photospheric abundances, since chemical fractionation processes in the outer atmospheric layers of stars may be present.
For the Sun the FIP (First Ionization Potential) effect is known, leading to an overabundance of low FIP ($\lesssim$~10\,eV) elements like Fe or Mg, 
whereas the high FIP elements Ne and O (that is used as reference element) are thought to remain
around photospheric values \citep[see e.g.][]{gei98}.

While the 'classical' solar Ne/O ratio from \cite{grsa} is Ne/O=0.18, the overall lower \cite{asplund} abundances result in a comparable value of Ne/O=0.15. 
However, the \cite{asplund} abundances require
a significant increase of the solar neon abundance by, depending on the assumed model and abundance uncertainties, a factor 
of $\sim$\,2.5 up to $\sim$\,4.0 \citep{ant05, bah05b}, i.e. Ne/O\,$\approx$\,0.4\,--\,0.6, could provide the missing opacity and would reconcile 
helioseismology with the new abundances derived by \cite{asplund}. Evidence for an increased neon abundance was presented by \cite{dra05},
who measured coronal Ne/O ratios in X-ray spectra of various nearby stars,
found a value of Ne/O=0.41 and 'solved' the problem by assuming a similar ratio for the Sun, thought noting that solar measurements indicate the lower values, i.e. Ne/O=0.15/0.18 to be correct. 
Actually their new neon abundance is still slightly below the required enhancement, but assuming other elemental
abundances to additionally increase jointly within allowed errors would provide a sufficient opacity increase.
Several objections to this solution were raised, especially from solar observers. A reassessment of solar coronal data from the {\it Solar Maximum Mission} led to an upper limit of
Ne/O=0.18$\pm0.04$ for active regions \citep{schmelz05} and an analysis of transition region lines
observed by the {\it SOHO} satellite suggested Ne/O=0.17$\pm0.05$ for the quiet Sun \citep{you05}, both confirming the 'classical'  solar Ne/O ratio.
Although sometimes higher Ne/O values were determined 
from X-ray spectra of energetic flares or in $\gamma$-ray production regions \citep[see e.g. Table 1, sup. data of][]{dra05},
solar measurements are overall consistent with a ratio of Ne/O$\approx 0.2\pm$0.05, as also noticed in \cite{dra05}. 
Furthermore, the stellar sample used by \cite{dra05} 
for their abundance analysis contains mostly active stars and binary
systems, which are not comparable to the Sun in terms of activity and are thought to exhibit a different chemical fractionation process.
This so-called inverse FIP effect \citep[see][for overall X-ray properties of stellar coronae]{gue04} is commonly observed in X-ray spectra of active stars.
It is possible that the inverse FIP effect modifies photospheric Ne/O ratios, leading to higher coronal values than in the photosphere.
Hence it is necessary to determine the Ne/O ratio in relatively inactive stars that are comparable to the Sun.
Since low activity stars are intrinsically X-ray faint, 
only a few nearby stars have been investigated and their Ne/O ratios were mostly found to be below the \cite{dra05} value but higher than the 'classical' solar one,
although measurement errors for individual stars were often quite large.

Two different methods have been used to investigate stellar coronal abundance ratios in general;
on one hand, by global spectral modeling
that usually includes the determination of key elemental abundances and some kind of emission measure distribution (EMD),
on the other hand, by using individual or a combination of emission lines that have a similar emissivity vs. temperature distributions and
thus the determined ratio is almost independent of the stellar EMD and the temperature dependence cancels out.
This latter method requires adequate and sufficiently strong lines for both elements in question, here Ne and O. Already \cite{act75} pointed out that the
emissivities of the \ion{O}{viii}~Ly$\alpha$ and the \ion{Ne}{ix} resonance line have a similar temperature dependence and are therefore 
suitable to determine the solar Ne/O ratio, which they determined to Ne/O=0.21$\pm0.07$. This method has been refined
by using linear combinations of suited lines to minimize temperature dependent residuals, specifically \cite{dra05} included the \ion{Ne}{x}~Ly$\alpha$ line 
and \cite{lie06} also the \ion{O}{vii} resonance line.
In general, inclusion of more lines, especially from different ionization stages, should minimize the temperature dependence, 
but adds measurement errors and restricts the method to detectors that spectrally resolve and cover all involved lines with sufficient sensitivity.

In this work we present a study of low to medium activity stars with log\,L$_{\rm X}$/L$_ {\rm bol}\lesssim -5$ and spectral types mid-F to mid-K,
for which high resolution X-ray spectra are available. All stars in our sample are main-sequence stars or only moderately evolved, i.e. they belong to luminosity classes IV\,--\,V.
We specifically determine coronal Ne/O abundance ratios for these stars by
using two linear combinations of Ne and O lines as well as global spectral fits and
check the results obtained from the different methods among each other and against  available literature values.
We further compare our results to those obtained for more active stars 
and investigate a possible abundance peculiarity of the Sun.

Our paper is structured as follows. In Sect.\,\ref{ana} we describe the applied methods and data used, in Sect.\,\ref{ress} we present our results,
check the applied methods and discuss implication of our findings in the context of solar and stellar coronal abundances and finally in Sect.\,\ref{con}
we summarize our findings and conclusions.

\section{Observations and data analysis}
\label{ana}

Our sample contains low to medium activity stars with spectral types mid-F to mid-K,
for which high-resolution X-ray spectra are available. In the next subsections we first describe the stellar sample and 
the observations, followed by a detailed description of the data analysis.

\subsection{Targets \& observations}

Our stellar sample consists of those stars from the {\it XMM-Newton}
monitoring program on coronal activity cycles, i.e. 61\,Cyg (K5+K7) \citep{hem06}, $\alpha$\,Cen (G2+K1) \citep{rob05,rob07cy} and HD\,81809 (G2+G9) \citep{fav04}, 
as well as the stars $\beta$\,Com (G0), $\epsilon$\,Eri (K2) and Procyon (F5).
The targets 61\,Cyg, $\alpha$\,Cen and HD\,81809 have been observed regularly for several years by {\it XMM-Newton} in short exposures separated half a year each.
These observations have resulted in roughly ten available exposures for each target, however
for the faintest target HD\,81809 we used only exposures taken during its X-ray brighter phases.
The high resolution spectra of HD\,81809 and 61\,Cyg are presented here for the first time and we reexamine $\alpha$\,Cen using a more
 extended dataset and analysis compared to \cite{lie06}. 
The components of the $\alpha$\,Cen system undergo strong, probably cyclic variability \citep{rob05,rob07cy}, therefore the contributions of the individual stars
to the flux of the system heavily depends on the phase of the observations.
This system is considered a prime test case since $\alpha$\,Cen~A is a close-by, nearly perfect solar twin and therefore should exhibit very similar properties. 

We further (re-)analyzed and re-examined {\it XMM-Newton} data of $\beta$\,Com, $\epsilon$\,Eri and Procyon 
and {\it Chandra}/LETGS data of $\alpha$\,Cen, $\epsilon$\,Eri and Procyon, which we retrieved from the archives.
In the context of stellar global coronal properties, the data of Procyon has been (partly) presented by \cite{raa02} and \cite{san04}, $\beta$\,Com is included in the
'Sun in time' sample \citep{tel05}, the LETGS spectra of $\alpha$\,Cen in \cite{raa03} and $\epsilon$\,Eri in \cite{wood06}. 
All observations used in this analysis are summarized in Table\,\ref{log}. 
We consider our dataset to constitute an overall representive of these stars and
include phases of activity common for these objects. 
Spatially resolved high resolution X-ray spectra of the binary systems are only available in the $\alpha$\,Cen/{\it Chandra} case, 
all other data represent the summed flux of both components of the respective system. The used
{\it XMM-Newton} data were taken with the EPIC (European Photon Imaging Camera) MOS detectors
and the RGS (Reflection Grating Spectrometer, 5\,--38\,\AA\,), all analyzed {\it Chandra} observations used the LETGS 
(Low Energy Transmission Grating Spectrometer, 2\,--170\,\AA\,) together with the HRC-S.

\begin{table} [ht!]
\setlength\tabcolsep{5pt}
\begin{center}
\caption{\label{log} Observing log of the X-ray data used and effective exposure time of the filtered high resolution data.}
\begin{tabular}{lllll}
\hline
Star & Mission & Year(s) &No Exp. & Obs.(ks)\\\hline
61 Cyg  &XMM & 2002-2007 & 11 & 103 \\
$\alpha$ Cen  & XMM & 2003-2007 & 9& 73 \\
$\alpha$ Cen & Chandra &1999 & 1 &79 \\
$\beta$ Com & XMM& 2003 & 1 & 41\\
$\epsilon$\,Eri& XMM & 2003 & 1 & 13 \\
$\epsilon$\,Eri & Chandra& 2001  & 1 & 105 \\
HD 81809 & XMM &2001-03, 2006-07 & 7 & 72  \\
Procyon & XMM & 2000, 2007 & 3 & 138 \\
Procyon & Chandra & 1999  & 2  & 139\\
\hline
\end{tabular}
\end{center}
\end{table}

\subsection{Data analysis}

The {\it XMM-Newton} data analysis was carried out with the Science Analysis System (SAS) version~7.0 \citep{sas}.
Standard selection criteria were applied to the data and
periods with enhanced background due to proton flares were discarded.
To increase the signal-to-noise-ratio of the RGS data
we further restricted the time intervals with acceptable background rates (CCD9 rate $<$\,0.5\,cts/s) for fainter sources like $\beta$\,Com and HD~81809 and
merged the individual observations for each targets using the tool '{\it rgscombine}'. 
The {\it Chandra} data was analyzed with standard tools of CIAO~3.4 \citep{ciao} we applied standard data processing and filtering and 
we used positive and negative first order spectra to measure line fluxes. 
The individual LETGS exposures provide sufficient data quality and are analyzed separately. 
To investigate global stellar X-ray properties, we use a combination of RGS/MOS data from the {\it XMM-Newton} observations,
for {\it Chandra} LETGS we use spectra in the 10\,--\,100\,\AA\, range.
For the determination of the Ne/O ratios only high resolution X-ray spectra with good SNR are taken into account.

Global spectral analysis was performed with XSPEC~V11.3 \citep{xspec} using
multi-temperature models (three temperature components are sufficient to describe our spectra)
with variable abundances as calculated with the APEC code \citep{apec}.  In our models,
temperatures, emission measure and abundances of Ne, O and Fe are treated as free parameters.
Depending on spectral quality, the abundances of other elements were also free parameters or set to solar values.
We also checked the results by fitting only those spectral ranges dominated by Ne and O lines,
 i.e. 11.0--14.0\,\AA\,, 18.0--23.0\,\AA\,, and additionally 85.0--100.0\,\AA\, for the LETGS.
We find that while the absolute abundances of Ne and O differ somewhat between the applied models, mainly due to the interdependence with the emission measure, 
the derived Ne/O ratio turns out to be a quite stable result.  

\begin{figure}[ht]
\includegraphics[width=92mm]{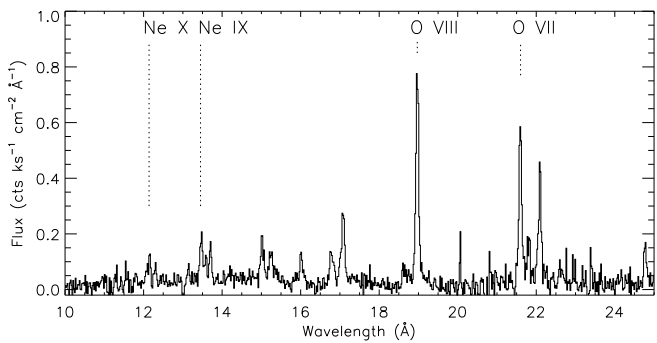}
\caption{\label{spec}High resolution X-ray spectrum of 61\,Cyg obtained with the RGS detectors; used spectral lines of neon and oxygen are labeled. }
\end{figure}

For line fitting purposes we use the CORA program \citep{cora}, using identical line width and assuming Lorentzian line shapes for 
all lines of a respective detector. This analysis uses total spectra, i.e. we determine one background\,(+continuum) level in the respective region around each line.
To derive the Ne/O ratios from linear combinations of emission lines
we especially measure the \ion{O}{viii}~Ly${\alpha}$ line (18.97~\AA) and the \ion{O}{vii} resonance line (21.6~\AA) as well as
the \ion{Ne}{x}~Ly${\alpha}$ line (12.13~\AA) and the \ion{Ne}{ix} resonance line (13.45~\AA).
Note that both Ly${\alpha}$ lines are actually unresolved doublets and that the strength of the resonance lines of \ion{O}{vii} and \ion{Ne}{ix} 
were derived from a fit to the respective triplet with all three lines included in the fit. As an example, we show
in Fig.\,\ref{spec} the X-ray spectrum of 61\,Cyg from the RGS detectors with the spectral lines of interest labeled. 
The spectrum consists of 11 co-added observations, background subtraction and flux conversion were done with the {\it 'rgsfluxer'} tool. 

\begin{figure}[ht]
\includegraphics[width=90mm]{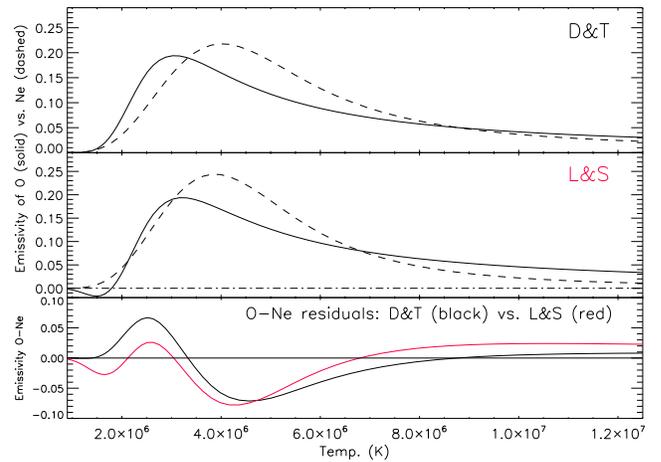}
\caption{\label{chianti} Theoretical emissivity curves for neon and oxygen as used by \cite{dra05}  {\it (upper panel)}
and \cite{lie06} {\it (middle panel)} and respective residuals {\it (lower panel)} vs. temperature. Note that D\&T used erg\,cm$^{3}$s$^{-1}$, while L\&S used photons\,cm$^{3}$s$^{-1}$, both in arbitrary units (see text).}
\end{figure}

\begin{figure}[ht]
\includegraphics[width=90mm]{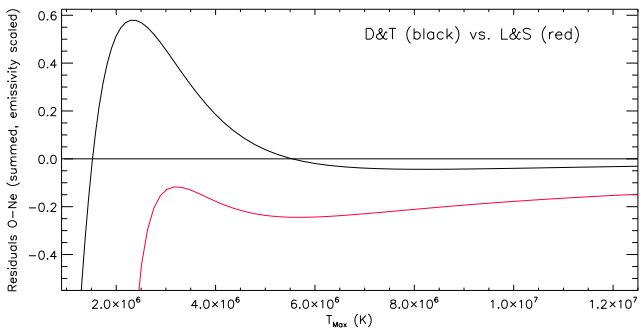}
\caption{\label{chiantires} Summed and scaled residuals of the emissivities (O\,--\,Ne) for the linear combinations used by D\&T and L\&S. 
Positive values correspond to an overprediction of oxygen, negative ones to an overprediction of neon (see text).}
\end{figure}

In the line ratio approach one uses linear combinations of these line fluxes to determine the Ne/O abundance ratio, however,
the specific linear combination can be differently chosen.
In Fig.\,\ref{chianti} we show the emissivity curves, i.e. G(T), of the respective linear combinations for Ne and O as used by \cite{dra05}\,(D\&T) and \cite{lie06}\,(L\&S)
and calculated with the CHIANTI~V\,5.2 code \citep{chi,anti}. Unfortunately the original weighting is based on energy flux in
D\&T (\ion{O}{viii} vs. $\ion{Ne}{ix}+0.15\times\ion{Ne}{x}$)
and on photon flux in L\&S ($0.67\times\ion{O}{viii}-0.17\times\ion{O}{vii}$ vs. $\ion{Ne}{ix}+0.02\times\ion{Ne}{x}$),
we therefore normalized the distributions for comparison. Note that the L\&S emissivity for O is unphysically negative at temperatures below 1.8\,MK, 
corresponding to an \ion{O}{viii}\,Ly$\alpha$/\ion{O}{vii}(r) energy flux ratio of $\sim 0.3$ or lower. As is apparent from Fig.\,\ref{chianti},
for both combinations the respective Ne and O curves are quite similar, but the residuals average out only over a very broad temperature range. 
It is important to keep in mind, that the temperature dependence of the residuals result in systematic
errors of the derived Ne/O ratios, which need not to be small in disadvantageous cases, e.g. for weakly active stars. In this case the exact value strongly depends 
on the used linear combination and the underlying stellar EMD. In order to illustrate this behavior, we show in Fig.\,\ref{chiantires} the summed residuals of the O--Ne emissivities scaled with the
summed, average emissivity vs. temperature; i.e. 

$\displaystyle \sum_{T=0}^{T_ {max}} \frac{G_{\rm Ne}(T)-G_{\rm O}(T)}{\frac{1}{2}\big(G_{\rm Ne}(T)+G_{\rm O}(T)\big)}$

\noindent
physically corresponding to the simplified case of cool star with a flat EMD over the respective temperature range. 
Neglecting very low temperatures where values are far outside the plotting range,
the summed residuals from D\&T are positive below $\sim$\,5\,MK and become slightly negative for higher temperatures, the L\&S values are negative for all temperatures.
In summary, the residuals in these emissivity ratios smooth out in stars with a sufficiently broad temperature distribution as noted by \cite{dra05};
however, this is not the case for stars with only cool or very dominant cool coronal plasma.

Here we investigate medium and low activity stars, whose coronal plasma temperatures are usually below 10\,MK and our sample stars are, as shown in Sect.\,\ref{ress}, clearly 
dominated by relatively cool plasma below 5\,MK during these observations.
In the spectra of weakly active stars \ion{Ne}{x} with a peak formation temperature of $\approx$~6\,MK is often virtually absent,
further its contribution to the emissivity of neon in the used linear combinations is secondary.
This makes the precise measurement of the \ion{Ne}{ix} resonance line a crucial point in this analysis that contributes greatly to the overall uncertainties. 
Further, contrary to the \ion{O}{vii} and \ion{O}{viii} lines, which are quite isolated in the X-ray spectra, \ion{Ne}{x} and
\ion{Ne}{ix} are found in a spectral region where several blends with Fe lines are present, which may influence the accurate determination of the neon abundance.
Since there are systematic, temperature dependent errors in the linear combinations, we
crosscheck these results with those from the multi-temperature models described above, which are based on a different analysis method and use a different data base.

\section{Results}
\label{ress}

Our sample stars differ in spectral type and level of activity as expressed by the L$_{\rm X}$/L$_{\rm bol}$ ratio, therefore we first present
global properties and give measured line strengths relevant for our study,
while the derived Ne/O ratios and a discussion of the results in the solar context will be presented in the following subsections.

\subsection{Global stellar properties}

We first determined the global X-ray properties of our sample stars from RGS and MOS spectra and present
in Table \ref{par} the respective X-ray luminosity, mean temperature and activity level 
together with basic stellar parameters adopted from the Simbad database.
For cyclic stars and unresolved binaries the X-ray data represent the average state of activity during the analyzed observations.
Bolometric luminosities were calculated from the apparent visual magnitudes,
the determined values of L$_{\rm bol}$ indicate that 
especially HD\,81809 and Procyon, which has luminosity class IV-V, are somewhat evolved and already above the main sequence.

\begin{table} [ht!]
\setlength\tabcolsep{3.5pt}
\begin{center}
\caption{\label{par} Basic parameters of our sample stars. L$_{\rm X}$ (0.2\,--\,3.0~keV) and X-ray flux ratios from RGS/MOS data.}
\begin{tabular}{lllllll}\hline\hline
Star & Type&V& Dist. &log\,L$_{\rm X}$ & T$_{\rm Av.}$& L$_{\rm X}$/L$_{\rm bol}$\\
 & & (mag)&(pc) &(erg/s) & (MK)& log \\\hline
61 Cyg & K5+K7& 5.2+6.0& 3.5 & 27.3 & 3.2&-5.6, -5.5 \\
$\alpha$\,Cen &G2+K1 & 0.0+1.3&1.3 & 27.1& 2.2 & -7.3, -6.2 \\
$\beta$\,Com &G0 &4.26 & 9.2 & 28.2& 3.4& -5.6\\
$\epsilon$\,Eri &K2 & 3.73& 3.2 & 28.2 & 3.8 & -4.9 \\
HD 81809 &(G2+G9)$^{*}$&5.8+6.8 &31.2 & 28.7 & 4.0& -5.6  \\
Procyon &F5+(WD)$^{*}$ &0.34 & 3.5 & 27.9& 1.9 & -6.5\\\hline
\end{tabular}
\end{center}
\begin{list}{}{}
\item[$^{*}$] Unresolved in X-rays.
\end{list}
\end{table}

\begin{figure}[ht]
\includegraphics[width=50mm,angle=-90]{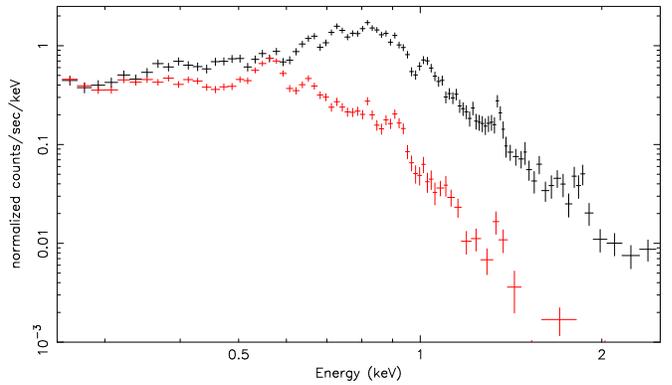}
\caption{\label{mos}X-ray spectra of $\epsilon$\,Eri (black) and Procyon (red) obtained with the MOS detector (1\,keV\,$\hat{=}$\,12.4\,\AA\,).}
\end{figure}

In Fig.\,\ref{mos} we show the MOS spectra
of the moderately active star $\epsilon$\,Eri and the weakly active stars Procyon. The star $\epsilon$\,Eri exhibits a considerably harder spectrum compared to Procyon,
indicating a significantly hotter corona.
We quote X-ray luminosities in the 0.2\,--\,3.0~keV band, i.e. where the MOS instruments provided useful data. 
The corresponding L$_{\rm X}$ in the 
{\it ROSAT} band (0.1\,--\,2.4~keV) is, depending on the spectral shape, roughly 20\,--\,40\% higher for our sample stars.
The binary systems 61\,Cyg and $\alpha$\,Cen are spatially resolved in the X-ray images obtained by the MOS detectors and
we give the average activity level of each component separately.
Both respective components are identified as variable X-ray emitters \citep{rob07cy}, with
 $\alpha$\,Cen~B (K1) and 61\,Cyg~A (K5) being the X-ray dominant components during our observations.
HD\,81809 is unresolved in all X-ray detectors and thus the main contributor to its X-ray emission is unknown. 

The derived average coronal temperatures of 2\,--\,4\,MK are typical for weakly to moderately active stars and their low X-ray luminosities, or 
alternatively L$_{\rm X}$/L$_{\rm bol}$, confirms this finding.
While our sample stars span about two orders of magnitude in activity,
really active stars have log~L$_{\rm X}$/L$_{\rm bol}\approx -3$ and are therefore a hundred times more active than our most active star.
The temperature of HD\,81809 appears quite high given its L$_{\rm X}$/L$_{\rm bol}$ ratio; this is
mainly due to an observation taken during its brightest phase when the source was also hot, however without exhibiting clear flare signatures \citep[see discussion in][]{fav04}.
In our global analysis we also investigated the abundances of other key elements and
we find indications of an enhancement of low FIP elements (Fe, Mg), resembling the solar FIP effect, in some of the less active stars like $\beta$\,Com and most clearly in $\alpha$\,Cen.
In contrast, the FIP effect is rather weak or even not present in the moderately active stars like 61\,Cyg and $\epsilon$\,Eri, but also in the more evolved low-activity star Procyon. 

\subsection{Neon and oxygen line fluxes}

We measured photon fluxes of the spectral lines of interest as described above from the respective high resolution spectra of our targets and listed their values in Table~\ref{res}.
For the $\alpha$\,Cen system we also investigated the data from the years 2003/04 separately to search for
possible changes in its Ne/O ratio. During these years $\alpha$\,Cen~A was in an X-ray bright phase and contributed most strongly to the
allover X-ray emission, on the other hand the overall activity level of the system was high. 
We also derive the \ion{O}{viii}({\small Ly$\alpha$})/\ion{O}{vii}(r) energy flux ratio for each dataset, a ratio that
is a good temperature proxy for the cooler X-ray emitting plasma and most sensitive in the 1.5\,--\,5\,MK temperature range. 
This ratio allows to crosscheck the global fitting results and is further used to investigate if the {\it Chandra} observations are performed at a comparable activity level for our targets.
We find that the \ion{O}{viii}/\ion{O}{vii} ratio correlates very well with the average temperature derived from the global spectral fitting of the MOS/RGS data.

\begin{table}[!ht]
\setlength\tabcolsep{3.5pt}
\caption{\label{res}Measured line fluxes in 10$^{-5}$ photons cm$^{-2}$ s$^{-1}$ from RGS or LETGS (C) and energy flux ratio of the \ion{O}{viii}({\tiny Ly$\alpha$}) to \ion{O}{vii}(r) line.}
\begin{center}{
\begin{tabular}{lllll|l}\hline\hline
Star & \ion{O}{vii}(r) & \ion{O}{viii} & \ion{Ne}{ix}(r) &\ion{Ne}{x}$^{\rm{*}}$&O8/O7(r)\\\hline
61 Cyg & 6.6$\pm$0.4& 9.1$\pm$0.5  & 2.0$\pm$0.3&1.6$\pm$0.2 & 1.57$\pm$0.13\\
$\alpha$\,Cen&33.5$\pm$1.3 & 27.0$\pm$1.2& 4.1$\pm$0.5& 1.6$\pm$0.2& 0.92$\pm$0.05 \\
$\alpha$\,Cen\,\tiny{03/04}&47.9$\pm$2.1 & 45.4$\pm$1.5& 5.5$\pm$0.8& 2.4$\pm$0.5 & 1.08$\pm$0.06\\
$\alpha$\,Cen A (C)& 9.2$\pm$1.0 & 3.2$\pm$0.5& 0.5$\pm$0.3& $\lesssim$ 0.2 & 0.40$\pm$0.07\\
$\alpha$\,Cen B (C)&11.5$\pm$1.0 & 6.3$\pm$0.6& 1.3$\pm$0.4& $\lesssim$ 0.2 & 0.62$\pm$0.08\\
$\beta$\,Com&3.8$\pm$0.6 & 5.9$\pm$0.6& 0.9$\pm$0.3& 0.6$\pm$0.3& 1.77$\pm$0.33\\
$\epsilon$\,Eri&44.1$\pm$2.6 & 78.0$\pm$3.0& 21.2$\pm$2.4& 10.5$\pm$1.4& 2.01$\pm$0.14\\
$\epsilon$\,Eri \tiny{(C)}&41.5$\pm$1.6 & 78.9$\pm$1.7& 18.8$\pm$0.9& 16.7$\pm$0.9& 2.16$\pm$0.10\\
HD 81809&1.0$\pm$0.3 & 2.3$\pm$0.3 &0.5$\pm$0.2& 0.5$\pm$0.2& 2.62$\pm$0.86\\
Procyon&35.6$\pm$1.1 & 22.9$\pm$0.9& 1.8$\pm$0.3& 0.5$\pm$0.2& 0.73$\pm$0.04\\
Procyon\,\tiny{(C)}&29.1$\pm$1.6 & 17.3$\pm$1.0& 2.5$\pm$0.5& 0.7$\pm$0.3& 0.68$\pm$0.05\\
Procyon\,\tiny{(C})&30.3$\pm$1.7 & 18.9$\pm$1.1& 1.4$\pm$0.3& $\lesssim$ 0.2& 0.71$\pm$0.06\\\hline
\end{tabular}}
\end{center}
\begin{list}{}{}
\item[$^{\rm{*}}$] Blended with \ion{Fe}{xvii}.
\end{list}
\end{table}

Comparing the {\it XMM-Newton} and {\it Chandra} measurements obtained for the same stars, we find nearly identical activity levels for Procyon 
and a slightly hotter but still comparable state in the LETGS data for $\epsilon$\,Eri. Completely different
values are found for $\alpha$\,Cen, which was apparently in a far cooler and X-ray fainter state during the {\it Chandra} observation, 
even when compared to the averaged RGS value from all  {\it XMM-Newton} observations. 
In comparison with the LETGS analysis by \cite{raa03}, we find similar values for the individual line fluxes, however our derived 
X-ray luminosities are significantly lower by a factor of $\sim$3. 
We find X-ray luminosities of roughly $3 \times 10^{26}$\,erg/s for $\alpha$\,Cen~A and $4 \times 10^{26}$\,erg/s for $\alpha$\,Cen~B, i.e. log\,L$_{\rm X}$=26.8 erg/s 
for the total system in the 0.2\,--\,3.0\,keV band, i.e. three times fainter than the derived L$_{\rm X}$ from the {\it XMM-Newton} measurements,
in line with the low fluxes measured in the emission lines and supported by the low \ion{O}{viii}({\small Ly$\alpha$})/\ion{O}{vii}(r) ratio.
In Procyon we find an apparent discrepancy by a factor of roughly two in the strength of the \ion{Ne}{ix} and \ion{Ne}{x} lines between 
the AO-0 (indicating higher fluxes) and AO-1 observations from {\it Chandra}/LETGS, although statistically only at the 2$\sigma$ level.
Since the RGS flux is intermediate between these measurements and the \ion{Ne}{vii} and \ion{Ne}{viii} lines around 88.1\,\AA\, are comparable in both LETGS observations, 
this might be a chance effect.

\subsection{Stellar Ne/O abundance ratios}

In order to obtain the true flux of the Ne and O lines, we have to take possible line blends into account. We identify
relevant blends from atomic line lists we inspected Chianti V5.2 and SPEX~V2.0, and deblended the lines when necessary with theoretical emissivities to ensure a uniform analysis.  
While no significant blends are present for the used O lines, the \ion{Ne}{x} line is blended with 
\ion{Fe}{xvii} at 12.12\AA\,, with both ions having similar peak formation temperatures of T$_{\rm Max}\approx$~5.5\,MK. Inspecting the respective line emissivities and 
assuming a solar abundance pattern the contamination of the \ion{Ne}{x} line is estimated to be around 20\%. 
The contamination is higher by a factor of up to three in those stars that show a 
strong FIP effect. Fortunately these stars tend to have intrinsically cool coronae and \ion{Ne}{x} is mostly weak or even undetected, allowing this simplified approach.
The \ion{Ne}{ix} triplet with T$_{\rm Max}=3.9$\,MK is again blended with several Fe lines, however the dominant blends are from \ion{Fe}{xix} (T$_{\rm Max}=8.1$\,MK) with its
strongest line being located at 13.52\,\AA\,. Even in our 'hottest' sample stars coronal plasma with temperatures around 8\,MK contributes only little
to the overall coronal emission and further this blend mainly affects the intercombination line at 13.55\,\AA\,, which is not used in this analysis.
We therefore expect an essentially negligible Fe contamination of the \ion{Ne}{ix} resonance line for our sample stars
and use its measured flux, whereas we reduce the measured \ion{Ne}{x} flux by 20\%.
The stellar Ne/O ratios from the linear combinations were then derived using these values and are given in Table~\ref{rat} together with the ratios
from global spectral modeling. The quoted errors are statistical errors due to spectral fitting and measurements of Ne and O lines only
and they do not take into account any systematic errors.

\begin{table}[!ht]
\caption{\label{rat}Derived Ne/O ratio for various methods (sorted by decreasing average activity).}
\begin{center}{
\begin{tabular}{llllll}\hline\hline
Star &D\&T & L\&S & global fit \\\hline
$\epsilon$\,Eri (RGS)&0.41$\pm$0.05 & 0.46$\pm$0.06 & 0.36$\pm$0.04 \\
$\epsilon$\,Eri (LETGS)&0.38$\pm$0.02 & 0.40$\pm$0.02 & 0.35$\pm$0.04\\
61 Cyg &0.34$\pm$0.05 &0.39$\pm$0.06 & 0.36$\pm$0.05\\
HD 81809 &0.36$\pm$0.15 & 0.35$\pm$0.13& 0.36$\pm$0.22\\
$\beta$\,Com (RGS) & 0.24$\pm$0.08 & 0.27$\pm$0.09 & 0.25$\pm$0.13\\
$\alpha$ Cen B& 0.29$\pm$0.07 & 0.54$\pm$0.16 & 0.26$\pm$0.16\\
$\alpha$ Cen (RGS)& 0.23$\pm$0.03& 0.32$\pm$0.04 & 0.25$\pm$0.03\\
$\alpha$ Cen (03/04-RGS)& 0.18$\pm$0.03 & 0.24$\pm$0.04 & 0.24$\pm$0.04\\
Procyon (RGS) & 0.12$\pm$0.02 & 0.19$\pm$0.03 & 0.21$\pm$0.03\\
Procyon (LETGS0) & 0.21$\pm$0.04 & 0.36$\pm$0.08 & 0.23$\pm$0.03\\
Procyon (LETGS1) & 0.10$\pm$0.03 & 0.18$\pm$0.04 & 0.21$\pm$0.03\\
$\alpha$ Cen A& 0.24$\pm$0.12 & 0.85$\pm$0.61 & 0.19$\pm$0.18\\\hline
\end{tabular}}
\end{center}
\end{table}

\begin{figure}[ht]
\includegraphics[width=92mm]{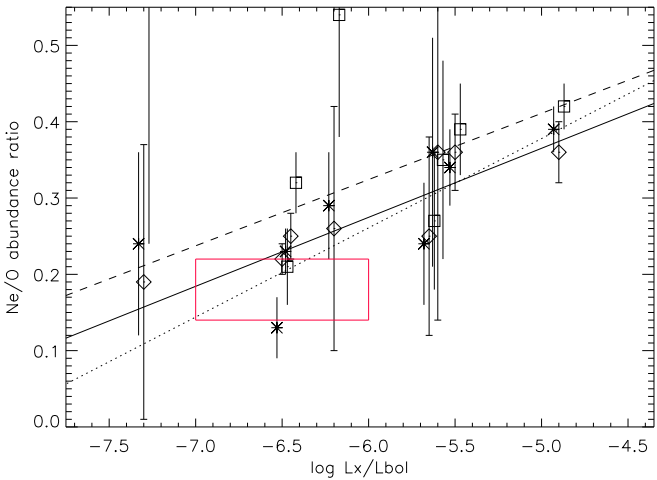}
\caption{\label{neo} Ne/O ratios of our sample stars derived with var. methods (global fit: diamonds/solid line; D\&T: asterisks/dotted line; L\&S: squares/dashed line) and the 'classical' Sun (red box).}
\end{figure}

Inspecting the results of the Ne/O ratios derived from the different methods, we find a quite good overall agreement for most sample stars.
Very similar values of the Ne/O ratio derived from the three methods are generally obtained for those stars
whose underlying EMD covers a sufficiently broad range of temperatures, provided that a well exposed spectrum is available.
However, this is hardly achieved by inactive stars and only the more active stars in our sample roughly fulfill this criterion. 
The most obvious discrepancies can thus be attributed to either systematic errors, that are expected from the theoretical calculations presented in Sect.\,\ref {ana},
or to cases where statistical errors are large.
There appears to be a tendency to an overprediction of Ne in the L\&S model for all sample stars and of 
O in the D\&T model for the lowest activity stars, exactly reflecting the theoretical expectations and hence pointing to the presence of systematic errors.
However, measurement errors and limited statistics prevent a more detailed comparison.
Therefore in many cases the true ratio should be in-between the results from two linear combinations and thus together they define a reasonable range of allowed values.
The Ne/O abundances derived from our global fits
indeed result mostly in intermediate values when compared to those from the line ratios, this is also represented by an intermediate regression curve shown in Fig.\,\ref{neo}.
In summary, both methods generally complement among each another, 
however the results derived from linear combinations of emission lines have to be treated with some caution when considering individual targets with unknown
underlying EMD. Nevertheless, all methods indicate clear correlation between Ne/O ratio and coronal activity.

In Fig.\,\ref{neo} we show the Ne/O ratio vs. stellar activity for our sample stars and the Sun as
derived from the three applied methods; slight offsets in log\,L$_{\rm X}$/L$_{\rm bol}$ among the methods are introduced for clarity. 
Our measurements clearly show a trend of lower Ne/O ratio towards less active stars, a finding that is independent of the applied method.
We also plotted the error weighted linear regression for each
method, whereas for stars with multiple observations the error weighted mean of the individual results is used.
Given our lack of any definite physical model,
the assumed linear dependence is the simplest possible fit over the considered parameter range. 
Although more complicated dependencies obviously cannot be ruled out,
a general decline of the Ne/O abundance with activity is present in all three regression curves.
Admittedly some of the results for individual stars have large statistical errors and our stellar sample is quite 
limited, however our results clearly suggest low Ne/O ratios for the weakly active stars. 
Also, the overplotted range of  'classical' solar values for activity and Ne/O ratio fits well into the respective parameter space derived from our low-activity stars.

Our results also agree reasonably well with available literature values (transformed to \cite{grsa} abundance scale) derived
from various multi-temperature EMD modeling, despite different applied methods in data analysis, underlying atomic data,
data sets or selection criteria.
Concerning the individual targets, the most active star in our sample, the K-dwarf $\epsilon$\,Eri, shows the highest ratio with Ne/O=0.35\,--\,0.4.
A similar range of values is present in the literature, e.g. Ne/O=0.36 \citep{wood06} or Ne/O=0.4 \citep{san04}.
For the group of slightly less active stars we measure Ne/O\,$\approx$\,0.35. The stars 61\,Cyg,  $\beta$\,Com and HD\,81809 
have very similar activity levels and there might to be a trend towards lower Ne/O ratios for stars of earlier spectral type at a 
given activity level. Especially the G0-star $\beta$\,Com is better described by a ratio around Ne/O=0.25, but given the errors and
the small sample size of only three stars, this has to be checked with larger and better observed stellar samples.
For the less active $\alpha$\,Cen system we find values around Ne/O=0.25. 
Additionally, in an analysis of selected phases of the  $\alpha$\,Cen data where the contribution of the lesser active component $\alpha$\,Cen~A to the RGS spectra is maximal,
indications of a decrease in the Ne/O ratio of the $\alpha$\,Cen system are present. This points to a lower Ne/O ratio in $\alpha$\,Cen~A compared to $\alpha$\,Cen~B.
\cite{raa03} derived values of Ne/O=0.18$\pm 0.07$~(A) and 0.24$\pm 0.09$~(B) for the individual components of the $\alpha$\,Cen system from the LETGS data, 
slightly lower than our values, but showing the same trend in the Ne/O ratio of the components as our results for the phase selected data. 
Also, the Ne/O=0.27 result from their line based analysis in \cite{lie06} fits into this picture, 
when keeping in mind that they used the RGS data of the $\alpha$\,Cen system from 2003\,--\,2005, i.e. intermediate in terms of
activity and contributing components between our 2003/04 dataset and the full data selection. However, given the measurement errors this has to be confirmed by further observations.
Procyon, the least active star in our sample beside $\alpha$\,Cen~A, again shows a very low Ne/O ratio for most of the data used. 
While \cite{san04} derived Ne/O=0.4 from the LETGS data (AO0+AO1), \cite{raa02} derived a value of Ne/O=0.22
for the RGS as well as the LETGS data (AO1). Some of these discrepancies are possibly due to the different analyzed datasets as discussed above. 
Although the even lower Ne/O ratios derived from the linear combination of D\&T probably have to be at least partially attributed to
systematic errors, values around Ne/O=0.2 are found repeatedly in various global analysis and linear combination models for Procyon. A value of Ne/O=0.2 also corresponds to the mean from
all used methods and datasets, whereas our averaged global analysis yields Ne/O=0.22 for Procyon.

In summary, for moderately active G- and K-dwarfs with activity levels of log\,L$_{\rm X}$/L$_{\rm bol}\approx -5\,...-5.5$ we measure abundance ratios of Ne/O=0.3\,--\,0.4.
While the Ne/O ratio further decreases with decreasing stellar activity, 
values around Ne/O=0.2\,--\,0.25 are found exclusively in the least active stars with activity levels around and below log\,L$_{\rm X}$/L$_{\rm bol}\approx -6.5$.
Additionally, none of these low activity stars shows a significantly higher ratio, therefore low Ne/O ratios seem to be common for stars with activity levels comparable to the Sun.
This finding is also supported by phase selected and spatially resolved analysis of the $\alpha$\,Cen system,
however measurement errors become large for these data.
A possible further dependence of spectral type, activity phase or other stellar parameters like luminosity class, binarity etc. might be present, but
cannot be investigated with the present data. 

We finally discuss the observed change of the Ne/O ratio in the context of more active stars. The ratio increases from Ne/O$\approx$0.2 up to Ne/O$\approx$0.4
when going from low to moderately activity stars, i.e. from log\,L$_{\rm X}$/L$_{\rm bol}\approx -7$ to log\,L$_{\rm X}$/L$_{\rm bol}\approx -5$.
Given these finding it is not surprising that \cite{dra05} found an average ratio of Ne/O=0.41 for a 
stellar sample dominated by stars or binary systems that are mostly ten to hundred times more active than the stars in our sample.
If anything, in the light of the inferred trend for the Ne/O ratio, their value appears rather low than high.
This further points to a different behavior when going to more extreme activity levels.  While
a linear dependence between Ne/O ratio and log\,L$_{\rm X}$/L$_{\rm bol}$ is a fair description for our stellar sample, we already noted that other models would also fit the data.
Taking their active main sequence stars and binary systems into account, we suspect the Ne/O ratio to saturate already at values around Ne/O\,$\approx$\,0.4\,--\,0.5. 
However, the detailed investigation of such a flattening or saturation of the Ne/O ratio towards very active stars is beyond the scope of this study, but deserves further attention.

\subsection{Discussion in the solar context}
\label{dis}

When putting the derived stellar Ne/O ratios in the context with the Sun, we first have to consider the solar activity level and the well known variations of the Sun's X-ray emission over the
course of the activity cycle. Further, most of the solar data refer to individual regions and the usually spatially resolved solar data has to be 
transformed into some kind of point-source like data that is available for stars. 
Using {\it Yohkoh}/SXT measurement taken from 1991\,--\,1995 and covering roughly a solar half cycle from maximum to minimum, \cite{act96} derived a conversion factor
between the observed X-ray index and the solar energy flux in the energy range 0.2\,--\,4.5\,keV, leading
to a solar activity range of log\,L$_{\rm X}$/L$_{\rm bol}\approx -7.9\,...-6.5$ and average coronal temperatures between 2\,--3\,MK. Another approach
based on {\it Yohkoh}/SXT data made by \cite{per00} resulted in a solar activity range over the cycle of log\,L$_{\rm X}$/L$_{\rm bol} \approx -7.2\,...-6.0$, while \cite{jud03} 
derived a slightly more active Sun with an activity range of log\,L$_{\rm X}$/L$_{\rm bol} \approx -6.9\,...-5.8$ from {\it SNOE}/SXP data. These literature values were transformed 
to our 0.2--\,3.0\,keV band by a 20\% reduction of the given 0.1\,--\,2.4\,keV flux. Note that even larger discrepancies by up to an order of magnitude are present between literature values
for the Sun's X-ray output \citep[see e.g. Table~2 in][]{jud03}, thus complicating a comparison with our sample stars. Averaged over recent measurements 
we find a solar X-ray luminosity range of $4\times10^{26}-4\times10^{27}$\,erg\,s$^{-1}$, i.e. an activity range over the cycle of
log\,L$_{\rm X}$/L$_{\rm bol}\approx -7\,...-6$. Hence the average X-ray activity level of the Sun appears to be comparably to those of the $\alpha$\,Cen system
and Procyon, while the Sun is significantly less active than the rest of the sample stars.
Given the above derived correlation between activity and Ne/O ratio, only these stars are considered further in our discussion on Ne/O ratios at solar-like activity levels.

Our low activity stars show ratios around Ne/O$\approx$0.2\,--\,0.25, i.e. close to the 'classical' solar ratio of  Ne/O=0.18$\pm$0.04 \citep{grsa}. This value is consistent with 
emission line measurements for the supergranule cells/network, representing the quiet Sun \citep{you05}, as well as for active regions \citep{schmelz05},
therefore the Ne/O ratio seems to be in general rather independent of the analyzed solar structure when overall properties are considered.
Spatially and temporally restricted analysis of X-ray and $\gamma$-ray lines from individual flares indicate a more diverse picture for local phenomena; 
they either point to an enhancement of low FIP elements \citep{mck92}, i.e. lower Ne/O ratios, a rather constant Ne/O ratio \citep{wid95}
or to an enhanced ratio in the range of Ne/O$\approx$0.25 \citep{ram95} up to Ne/O$\approx$0.3 \citep{schmelz93}. 
While these events may not be overall representive 
of the atmospheric layers discussed here, the chemical fractionation also seems to depend on the particular properties of the observed event. 
Further, fairly constant Ne and O abundances were also found in the composition of the solar wind or flare particles obtained by various instruments \citep{gei98},
again referring to a rather average solar property.
Therefore all present evidence points to 'unified' abundance ratio of Ne/O$\approx 0.2\pm$0.05 in the outer atmospheric layers of the Sun and other weakly active stars.  
We note, that Ne/O=0.15 from the revised abundances
of \cite{asplund} does not fit equally well, but is also consistent with the stellar data given the present uncertainties. However, in this case the
solar interior problem remains. 

A significantly higher neon abundance around Ne/O$\approx$\,0.5 as proposed for reconciliation between helioseismology and the revised abundances
can be ruled out for the coronae of weakly active stars, consistent with the findings for the Sun's outer layers.
A higher photospheric neon abundance is of course not in conflict with our data, however such a
scenario would require some unknown but universal fractionation process that essentially decouples the
photospheric neon abundance from the coronal one in the Sun and other low activity stars without significantly influencing oxygen at the same time. 
This hypothesized mechanism would have to operate, despite both elements being high FIP elements, have similar properties and ionization times in solar fractionation modeling,
exhibit a roughly constant ratio in various solar surface features of different activity level or show similar properties in solar wind data \citep{vst89, gei98}. 
We note that chemical fractionation models for different solar layers and structures are still under debate \citep[e.g.][]{mck98}, however
we feel that such a scenario seems to be quite unlikely and suppose that further revised solar modeling will settle the matter. 

Thus we cannot resolve the discrepancy between recent modeling of solar abundances and helioseismology with our stellar Ne/O abundance measurements
and no other element than is easily at hand to provide the missing opacity, due to the fact that the abundances of other important elements can be constrained by photospheric measurements.
However, at least solar coronal properties appear to fit very well into the X-ray picture derived for low activity stars with similar spectral type and evolutionary phase.
Therefore the Sun seems to be in no way exceptional or peculiar and behaves just like a typical, weakly active, middle aged star.

\section{Summary and conclusions}
\label{con}

   \begin{enumerate}
      \item We have determined the coronal Ne/O abundance ratio in a sample of low and moderately active stars. For stars in the activity range log\,L$_{\rm X}$/L$_{\rm bol}\approx -5\,...-7$
we find a trend of decreasing Ne/O ratio with decreasing activity level. 
The Ne/O ratio decreases from values of Ne/O\,$\approx$\,0.4 for moderately active stars down to Ne/O\,$\approx$\,0.2\,--\,0.25 for low activity stars.
The decrease is measured independently of the chosen analysis method, i.e. by global spectral modeling and two linear combinations of strong emission lines. 

\item The Sun as a low activity star fits very well into the picture derived from the stellar X-ray data. A low Ne/O ratio around the  'classical' solar value of Ne/O=0.18 might be
rather typical for solar-like stars with comparable activity level. A significant higher neon abundance, i.e. a ratio of Ne/O=0.4\,--\,0.6, 
as proposed to solve the solar interior problem can be virtually ruled out for the coronal composition. 
It also appears unlikely for the photospheric composition given the present knowledge and observational data
on chemical fractionation processes in weakly active stars.

\item We consider the main findings of this study to be robust, however the presently available data is insufficient to determine the detailed characteristics of the decline. 
While a linear dependence between Ne/O ratio and activity level describes our data well, some kind of flattening or saturation is expected for very active stars.
Possible dependences of abundance fractionation on other stellar parameters beside activity also remain unclear. However, further X-ray
observations of nearby low activity stars may successfully address these problems and tighten their re-unification with the Sun.

   \end{enumerate}

\begin{acknowledgements}
This work is based on observations obtained with XMM-Newton, an ESA science
mission with instruments and contributions directly funded by ESA Member States and NASA and it has made use of data obtained from the Chandra Data Archive.
J.R. acknowledges support from DLR under 50OR0105.

\end{acknowledgements}

\bibliographystyle{aa}
\bibliography{/data/hspc44/stch320/Pubs/jansbib}

\end{document}